\documentclass{article}
\usepackage{graphicx} 
\usepackage[numbers]{natbib}
\usepackage{amsmath}
\usepackage{amssymb}
\usepackage[margin=1in]{geometry}
\usepackage{xcolor}
\usepackage{booktabs}
\definecolor{myboxcolor}{RGB}{101, 117, 248}
\usepackage{hyperref}

\title{Flow Matching for Convective-Scale Precipitation Downscaling}
\author{Tom Wetherell \\ \small\textit{tomas.wetherell@metoffice.gov.uk}}\date{May 22, 2026}
\begin{document}

\maketitle

\begin{abstract}
Generative machine learning is an increasingly important complement to dynamical downscaling for producing high-resolution precipitation projections, with diffusion models currently the leading approach. Flow matching is a related generative framework that has recently achieved strong results across image, video and other domains, and shown early promise for downscaling. We train a flow matching model to map daily precipitation from 8 km to 2 km over a convective-scale domain centred on Singapore, and benchmark it against CPMGEM, a score-based diffusion model. Flow matching achieves consistently better spatial skill: higher fractions skill score at every precipitation threshold and neighbourhood scale tested, and tighter structure and amplitude components of the SAL score with comparable location skill. However, flow matching underestimates the upper tail of the precipitation distribution, resulting in a dry bias in the climatological mean. These results suggest that flow matching is a competitive generative framework for convective-scale precipitation downscaling, particularly well suited to capturing spatial structure.
\end{abstract} \hspace{10pt}

\section{Introduction}

High-resolution climate information is crucial for disaster mitigation, risk assessment, and climate change adaptation planning. Global climate models (GCMs) simulate the Earth system on a global grid, and are the primary tool for projecting future climate.  Due to their coarse spatial resolution (typically 25-150 km), GCMs are unable to resolve complex topography and mesoscale processes such as convective storms -- limiting their applicability for assessing local impacts and extreme events \citep{maraun2010precipitation}.

Regional climate models (RCMs) are the established tool for bridging this resolution gap. RCMs dynamically downscale GCM output over a limited-area domain, taking GCM fields as lateral boundary conditions and numerically simulating the atmosphere at finer resolution (typically 2-12 km). This approach effectively resolves local topography and small-scale dynamics \citep{kendon2012realism, needforrcms}, but is computationally expensive -- often limiting the range of climate scenarios and time periods that are downscaled, as well as the number of ensemble members generated. 

Machine learning (ML) downscaling has emerged as a complementary, computationally inexpensive approach \citep{kendon2025potential}. By training on paired low- and high-resolution data (for example, GCM-RCM pairs), ML models can learn a mapping from coarse to high-resolution fields. Once trained, inference is orders of magnitude cheaper than running an RCM, which opens up applications infeasible with dynamical downscaling alone: filling temporal gaps in existing high-resolution archives; generating large ensembles to better characterise extreme events; and downscaling GCMs for which no dynamical downscaling exists.

Deterministic convolutional architectures such as CNNs and U-Nets were among the first ML approaches applied to climate downscaling \citep{deepesd, doury2023regional, doury2024suitability}. They have shown success in reproducing the observed climate at a distributional level. However, their outputs are typically overly smooth -- lacking fine-scale spatial detail and under-representing extremes \citep{wardleikis2025intercomparisongenerativemachinelearning}. This is a particular issue for precipitation, which is spatially heterogeneous and heavy-tailed. Further, without modifications to enable probabilistic inference, they are unable to generate multiple ensemble members. 

Diffusion models represent the current state of the art in ML downscaling. \citet{cpmgem2024} applies a score-matching diffusion model to learn a mapping from coarse atmospheric fields to km-scale precipitation over the UK. \citet{corrdiff} introduces a two-step approach; a deterministic U-Net is first used to predict the conditional mean, and a diffusion model is used to add spatial detail by predicting the residual. Diffusion models have been shown to better represent fine-scale spatial detail and, being generative, are able to sample multiple plausible high-resolution fields conditioned on the same coarse input. This stochasticity allows them to be used to efficiently generate large ensembles, enabling scientists to better characterise uncertainty and quantify the likelihood of extreme events. 

Flow matching \citep{fm_lipman, fm_albergo} is a related generative framework that has recently achieved state-of-the-art results across image, video, protein and robotics domains, and shown early promise for downscaling \citep{sfm}. In this work, we train a flow matching model to map daily precipitation from 8 km to 2 km over a convective-scale domain centred on Singapore, and benchmark it against CPMGEM \citep{cpmgem2024}, a state-of-the-art score-based diffusion model. Flow matching achieves consistently stronger spatial skill --- higher fractions skill score at every threshold and neighbourhood scale tested, and tighter structure and amplitude components of SAL with comparable location skill --- though it under-represents the upper tail of the precipitation distribution relative to CPMGEM. Together these results suggest that flow matching is a competitive generative framework for convective-scale precipitation downscaling, particularly well suited to capturing spatial structure.

\section{Flow Matching for Downscaling}

The goal of generative modelling is to generate samples from an underlying data distribution, typically conditioned on auxiliary inputs. In the case of downscaling, we condition the generation on coarse-resolution atmospheric field(s) and aim to sample from the conditional distribution of high-resolution field(s) (for example, high resolution precipitation) consistent with the coarse-resolution input. The stochasticity of generative models makes them a natural fit for downscaling, as many high-resolution fields may be consistent with any given coarse input. 

Flow matching is one such generative framework, which learns to transport samples from a source distribution to a target distribution. Given a source distribution $p_0$ and a target distribution $p_1$, we define an interpolant \citep{fm_albergo}
\begin{equation}
x_t = \alpha_t x_1 + \beta_t x_0, \quad t \in [0, 1],
\label{eq:interpolant}
\end{equation}
where $x_0 \sim p_0$, $x_1 \sim p_1$, and  $\alpha_t,  \beta_t : [0, 1] \rightarrow \mathbb{R}$ are scalar functions satisfying $\alpha_0 = 0$, $\beta_0 = 1$ and $\alpha_1 = 1$, $\beta_1 = 0$. 
The interpolant induces a probability path $p_t$ with $p_{t=0} = p_0$ and $p_{t=1} = p_1$. \citet{fm_lipman} show that there exists a time-dependent velocity field $u_t$ generating this path, in the sense that solutions of the ordinary differential equation $\text{d}x_t/\text{d}t = u_t(x_t)$ initialised at $x_0 \sim p_0$ satisfy $x_t \sim p_t$ for all $t \in [0, 1]$. In particular, $x_1 \sim p_1$, and so the ODE induced by $u_t$ transports samples from $p_0$ to $p_1$.
 
We parameterise the velocity field with a neural network $u_t^{\theta}$. While $u_t$ itself is intractable, \citet{fm_lipman} show that $u_t^{\theta}$ can be trained to approximate $u_t$ by regressing against the conditional velocity field $\dot \alpha_t x_1 + \dot \beta_t x_0$ associated with each endpoint pair $(x_0, x_1)$, via the conditional flow matching (CFM) loss 
\begin{equation}
\mathcal{L}_{\mathrm{CFM}}(\theta)  
=  
\mathbb{E}_{t \sim \mathcal{U}(0,1),\, x_0 \sim p_0,\, x_1 \sim p_1}  
[\|  
u_t^\theta(x_t)  
-  
(  
\dot{\alpha}_t x_1 + \dot{\beta}_t x_0  
)  
\|^2].
\label{eq:cfm_loss}
\end{equation} 

In the conditional setting, the velocity network is additionally conditioned on $y$, i.e., $u_t^\theta(x_t, y)$, and the expectation in the CFM loss is taken over paired samples $(x_1, y)$ from the joint distribution $p(x_1, y)$. The source $x_0 \sim p_0$ is sampled independently of $(x_1, y)$. In our downscaling setting, $y$ is the coarse-resolution atmospheric fields and $x_1$ the corresponding high-resolution precipitation field. 

At inference, given a coarse-resolution conditioning input $y$, samples are generated by drawing $x_0 \sim p_0$ and numerically integrating the ODE induced by $u_t^\theta$ from $t=0$ to $t=1$, yielding a sample $x_1 \sim p(\cdot | y)$.

\section{Experimental Setup}

\subsection{Data}

We use data from Singapore's Third National Climate Change Study (V3) \citep{ccrs2024v3science}. V3 provides high-resolution climate projections over Southeast Asia by applying a two-stage dynamical downscaling approach; a collection of six CMIP6 GCMs (75-250 km resolution) are downscaled to 8 km over a large region encompassing most of Southeast Asia, and subsequently to 2 km over a smaller inner domain covering Singapore, the Malay Peninsula and Sumatra.  The SINGV-RCM regional climate model \citep{dipankar2020singv} was used for both downscaling stages. In this work, we use the 8 km and 2 km RCM simulations as the input and target of our downscaling task, respectively. The model is therefore trained to emulate the second stage of the V3 downscaling, using the imperfect model framework \citep{van2023deep}.

From V3 we use a single GCM-scenario pair -- UKESM1-0-LL driven by SSP5-8.5 -- and restrict to a $128 \times 128$ grid-cell cutout of the 2 km inner domain, centred on Singapore. The corresponding 8 km predictor region is a $32 \times 32$ cutout over the same geographic area, upsampled to $128 \times 128$ to match the target resolution. This work represents a first step towards a patch-based approach to downscale the full $960 \times 960$ 2 km domain. 

The predictor variables used are temperature, specific humidity and the eastward and northward components of wind, each at 200, 500, 700 and 850 hPa, together with sea-level pressure, taken at daily frequency from the 8 km simulation. We additionally include orography and a land-sea mask at their native 2 km resolution. The atmospheric predictor variables are normalised with z-score normalisation; orography is square-rooted and then rescaled to $[0, 1]$. The target is daily-mean precipitation at 2 km resolution, transformed by taking the square root and linearly rescaling to $[-1, 1]$.  All normalisation statistics are computed over the training period. 

The data is split temporally into training, validation and test sets. The training set comprises 1995-2014 and 2080-2099, exposing the model to two distinct climates spanning recent-past and end-of-century conditions. The validation set is 2040-2049, and the test set is 2050-2059. 

\subsection{Baseline}

We compare against CPMGEM \citep{cpmgem2024}, a diffusion-based downscaling model built on the score-based generative modelling framework of \citet{song2021scorebasedgenerativemodelingstochastic}. The network architecture, normalisation, and training hyperparameters follow the original implementation. We adapt the spatial dimensions of the score network to the $128 \times 128$ grid, and reduce the number of Euler sampling steps from 1000 to 300 (resulting in 300 network function evaluations per sample), following hyperparameter tuning on the validation set. 

\subsection{Flow Matching Implementation}

We use a standard Gaussian source distribution $p_0 = \mathcal{N}(0, I)$ and set $\alpha_t = t$, $\beta_t = 1 - t$, yielding the linear interpolant $x_t = (1-t)x_0 + t x_1$.

The velocity field is parameterised with an ADM U-Net \citep{dhariwal2021diffusionmodelsbeatgans} operating at $128 \times 128$ resolution. The network has 64 base channels with multipliers $[1, 2, 3, 4]$, two residual blocks per resolution level, and self-attention at the $16 \times 16$ feature map. Conditioning is provided by concatenating the upsampled coarse-resolution predictors, the static fields, and the interpolant $x_t$ along the channel dimension. Additionally, to allow the network to condition on absolute spatial location, we include 4 channels for sinusoidal positional embedding (sine and cosine of a half-cycle wave along each spatial dimension). The interpolation time $t \in [0, 1]$ is encoded with a Fourier embedding \citep{tancik2020fourierfeaturesletnetworks}. The U-Net architecture comprises approximately 23 million learnable parameters. 

The network is trained by minimising the conditional flow matching loss \eqref{eq:cfm_loss} on the training set. We use the AdamW optimiser with a learning rate of $1.0 \times 10^{-4}$, $\beta_1=0.9$ and $\beta_2=0.999$. An exponential moving average (EMA) of the network weights is maintained throughout training with decay rate 0.999, and used for inference. Training was performed on a single A100, and took 130 minutes for 153 epochs.

At inference, given coarse-resolution conditioning, we draw a source sample $x_0 \sim \mathcal{N}(0, I)$ and integrate the ODE induced by the learnt velocity field from $t=0$ to $t=1$ using Heun's method with 50 steps (corresponding to 100 network function evaluations per sample). We use a non-uniform time-step schedule $t_i = 1 - (1 - i/n)^2$ for $i = 0, ..., n$, which takes smaller steps near $t=1$ as the trajectory approaches the data distribution. Ensembles of 32 samples per conditioning are generated by varying the source draw $x_0$. 

\section{Results}

We assess model performance using the continuous ranked probability score (CRPS) as our headline metric. CRPS is a standard metric for evaluating probabilistic forecasts, comparing the full predicted distribution to the observed value. It accounts for both the location and spread of the forecast, with lower values indicating a closer match to the observations. Table~\ref{tab:crps} reports CRPS for both models. 

\begin{table}[h]
\centering
\begin{tabular}{lc}
\toprule
Model & CRPS \\
\midrule
Flow matching & \textbf{4.59} \\
CPMGEM & 4.70 \\
\bottomrule
\end{tabular}
\caption{CRPS over the test period (2050--2059), averaged over time and spatial dimensions. Units are mm/day. Computed from 32-member ensembles. Lower is better.}
\label{tab:crps}
\end{table}

Flow matching achieves a lower CRPS than CPMGEM at one-third the sampling cost (100 vs 300 network evaluations per sample). We next decompose performance along two axes: Section~\ref{sec:spatial} assesses spatial skill via fractions skill score and SAL, and Section~\ref{sec:intensity} examines intensity calibration.

\subsection{Spatial skill}
\label{sec:spatial}

We assess spatial skill using two complementary metrics: the fractions skill score (FSS) \citep{fsslean2008}, a neighbourhood-based metric, and the structure-amplitude-location (SAL) score \citep{wernlisal2008}, which decomposes performance into object-level components.

\begin{figure}[!htbp]
    \centering
    \includegraphics[width=0.8\textwidth]{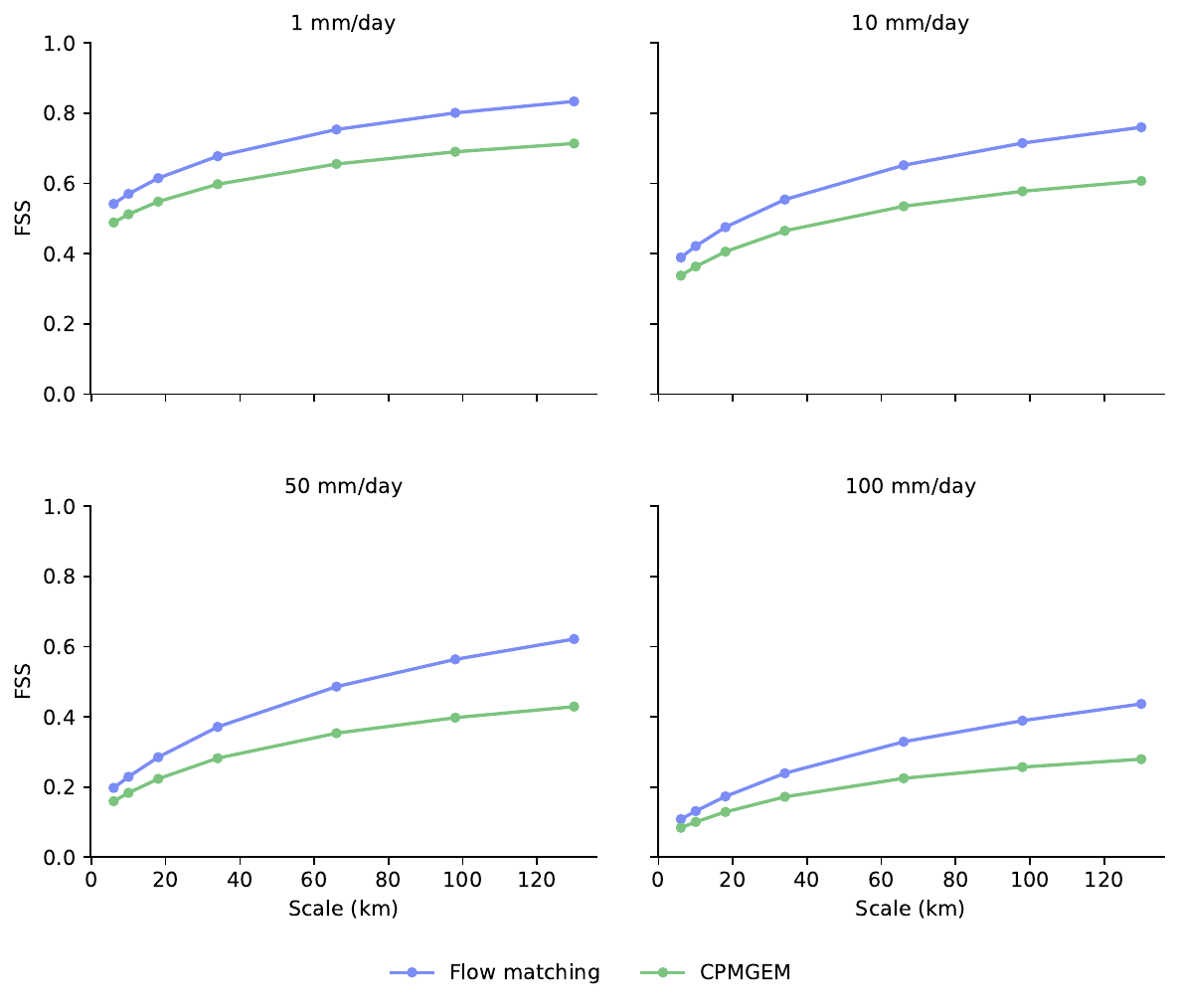}
    \caption{Fractions skill score as a function of neighbourhood scale for flow matching (blue) and CPMGEM (green), at four precipitation thresholds. Computed jointly across the ensemble and test period. Higher is better.}
    \label{fig:fss}
\end{figure}

\paragraph{Fractions skill score.}
FSS measures the agreement between binary precipitation masks (defined by a threshold) over neighbourhoods of varying size. FSS is computed using the \texttt{scores} package \citep{Leeuwenburg_scores_A_Python_2024, leeuwenburg_2026_18638494} with default aggregation across the ensemble and test period. We report scores at four thresholds spanning light to extreme precipitation -- 1, 10, 50 and 100 mm/day -- and at neighbourhood widths of 3, 5, 9, 17, 33, 49, and 65 grid cells (corresponding to 6--130 km).
Figure~\ref{fig:fss} shows that flow matching achieves higher FSS than CPMGEM at every threshold and every scale tested. Notably, the advantage persists at the 100 mm/day threshold, where one might expect the reduced upper-tail intensity of the flow matching samples (Section~\ref{sec:intensity}) to degrade scores. This suggests that across the range of thresholds examined, flow matching better represents the spatial distribution of precipitation at a given magnitude; intensity calibration issues are confined to the far upper tail.

\paragraph{Structure--Amplitude--Location.}
The SAL framework \citep{wernlisal2008} provides an object-based assessment of precipitation forecasts. Each field is segmented into objects by thresholding at a fraction of its maximum; the forecast is then scored along three axes: structure ($S$), comparing the size and shape of objects; amplitude ($A$), comparing the total precipitation amount; and location ($L$), combining the displacement of the field's centre of mass with the dispersion of objects about it. Perfect agreement is $S = A = L = 0$. $S$ and $A$ are signed and bounded in $[-2, 2]$; $L$ is non-negative, bounded in $[0, 2]$.

We compute SAL using \texttt{pysteps} \citep{pysteps} with the default object-detection parameters, independently for each (ensemble member, day) pair over the test period. Pairs in which no objects are detected in either field yield undefined $L$ and are excluded. To focus the evaluation on heavy-precipitation events, we restrict to days where the target field contains at least one grid cell exceeding 100 mm/day (2,266 of 3,652 test days).

\begin{figure}[!htbp]
    \centering
    \includegraphics[width=0.9\textwidth]{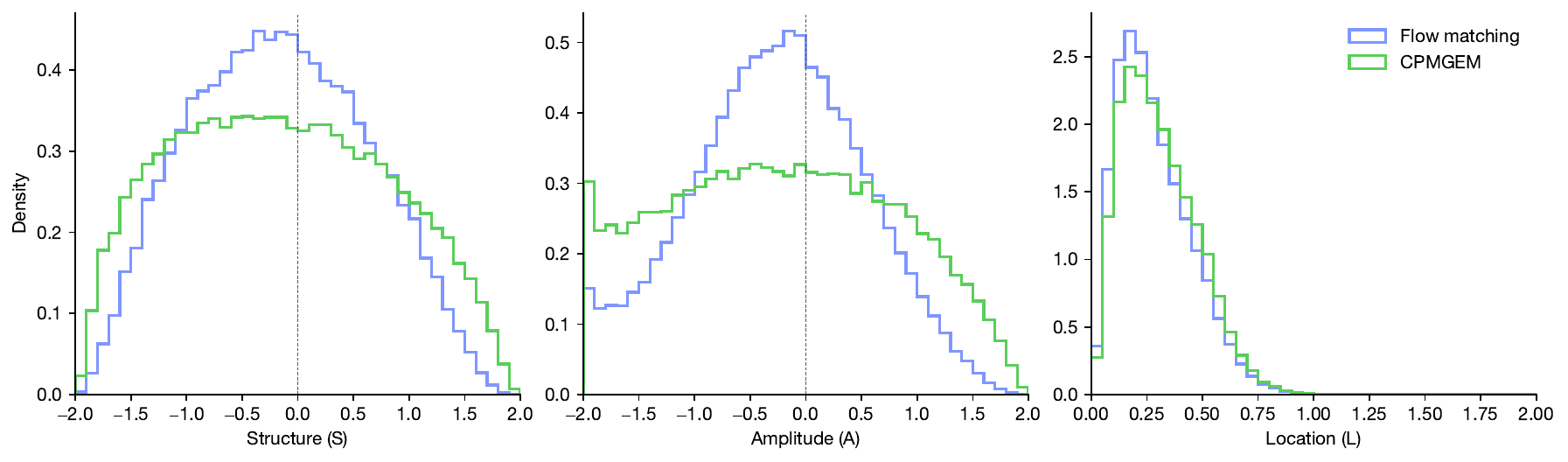}
    \caption{Distributions of SAL components --- structure ($S$), amplitude ($A$), and location ($L$) --- for flow matching (blue) and CPMGEM (green). $S$ and $A$ are signed and zero for perfect agreement; $L$ is non-negative with zero indicating perfect agreement. Distributions are pooled across all 32 ensemble members and all days in the test period (2050--2059) where the target contains at least one grid cell exceeding 100 mm/day; pairs with no detected objects in either field are excluded.}
    \label{fig:sal}
\end{figure}

Figure~\ref{fig:sal} compares the resulting distributions for the two models. Flow matching produces noticeably tighter distributions of $S$ and $A$, both centred close to zero --- indicating more consistent agreement on object size/shape and on total daily precipitation amount. The $L$ distributions are similar between the two models, indicating comparable skill at placing the overall centre of mass of precipitation. Combined with the FSS results, this suggests that flow matching's spatial advantage comes from better structure and local coverage rather than overall displacement. 

\subsection{Intensity calibration}
\label{sec:intensity}

\begin{figure}[!htbp]
    \centering
    \includegraphics[width=0.8\textwidth]{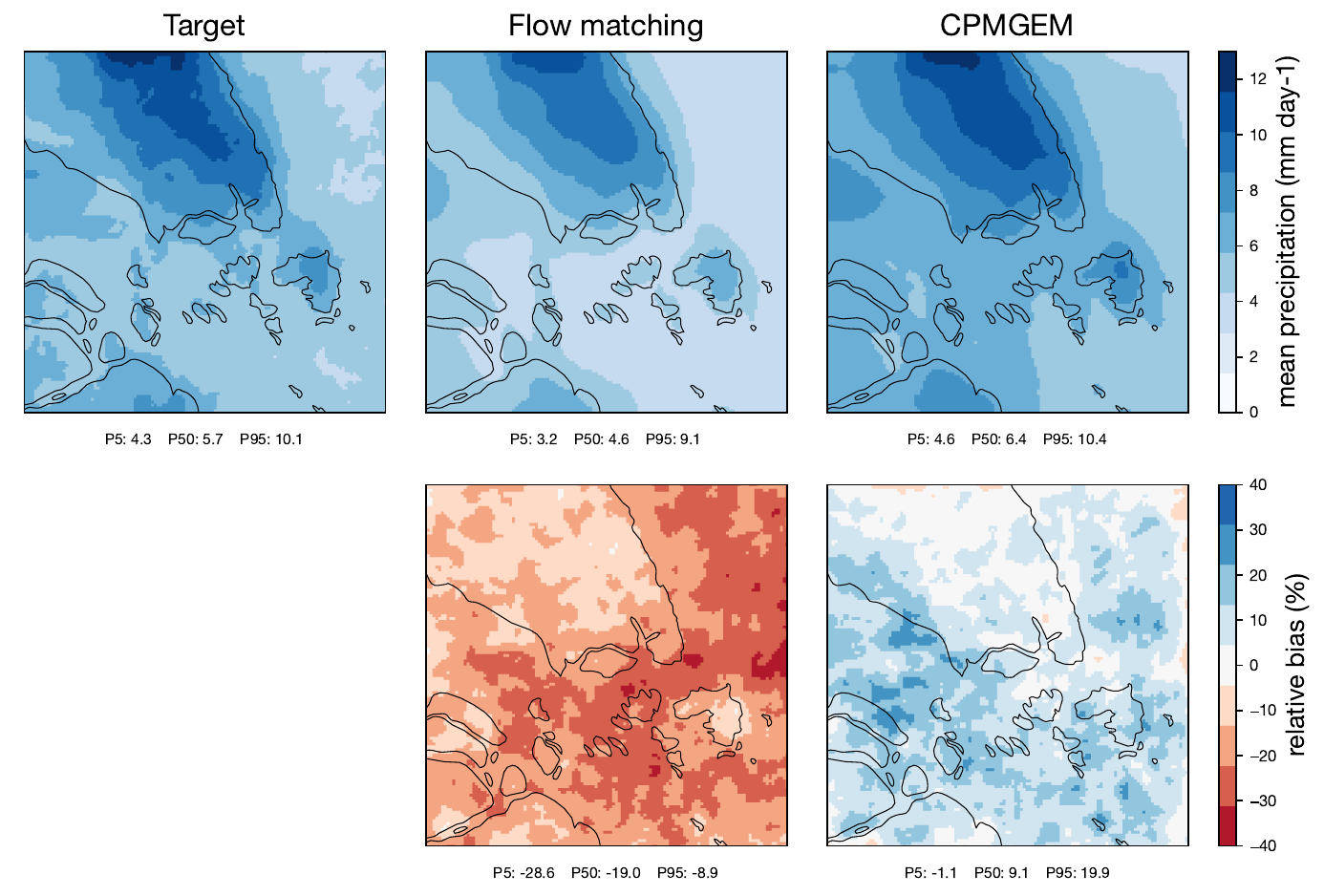}
    \caption{Daily-mean precipitation climatology over the test period (2050--2059). Top row: time-mean precipitation for the RCM target (left), flow matching (centre) and CPMGEM (right); model fields are additionally averaged over the 32-member ensemble. Bottom row: relative bias of each model against the target, $(\overline{x}_{\mathrm{model}} - \overline{x}_{\mathrm{target}}) / \overline{x}_{\mathrm{target}} \times 100\%$. Annotations below each panel report the spatial 5th, 50th and 95th percentiles (P5, P50, P95) of the displayed field.}
    \label{fig:climatology}
\end{figure}

Figure~\ref{fig:climatology} shows the time-mean precipitation field and the relative bias of each model against the target over the test period. Flow matching exhibits a systematic dry bias across the entire
domain -- particularly over the sea. CPMGEM is closer to the target on the time mean, with a small positive  bias.

\begin{figure}[!htbp]
    \centering
    \includegraphics[width=0.8\textwidth]{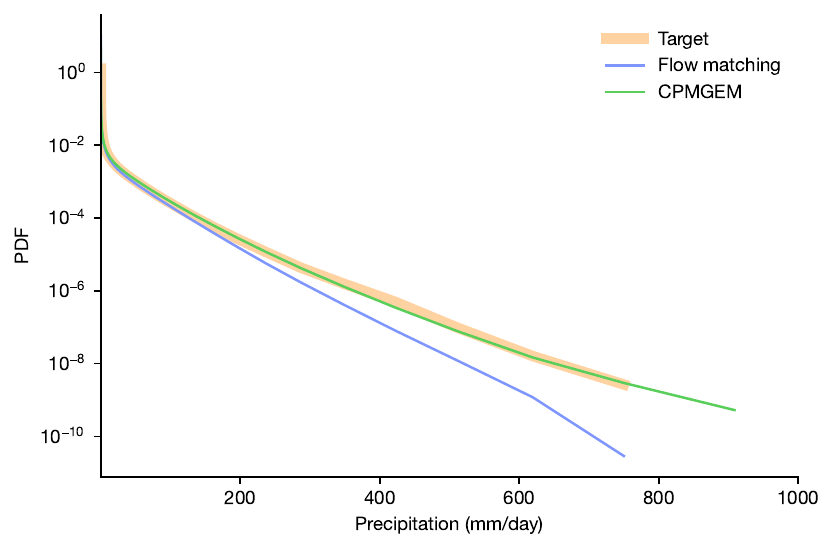}
    \caption{Probability density function of daily precipitation (mm/day) for the target (orange), flow matching (blue) and CPMGEM (green). Densities are estimated from histograms over 60 logarithmically spaced bins between 0.01 and 1000 mm/day and shown on a log y-axis to highlight the heavy upper tail.}
    \label{fig:distribution}
\end{figure}

The precipitation distributions (Figure~\ref{fig:distribution}) clarify the
origin of the bias. Both models track the target closely at low to
moderate intensities, but in the upper tail the flow matching model produces
heavy precipitation values far less frequently than the target, while
CPMGEM tracks the target into the deep tail. The under-representation of the upper tail is consistent with the dry bias seen in the time mean.

\section{Discussion}

This work demonstrates that flow matching is a competitive generative framework for convective-scale precipitation downscaling. On a 2 km Singapore domain, our flow matching model shows superior spatial skill to the diffusion baseline -- as measured by FSS and SAL scores. A notable weakness is underestimation of the upper tail of the precipitation distribution, which propagates to a dry bias in the climatological mean. 

An exciting next step is to extend this single-patch setup to a patch-based scheme covering the full $960 \times 960$ 2 km Western Maritime Continent domain. This would demonstrate the model's potential for filling temporal gaps in the V3 archive at a fraction of the cost of dynamical downscaling. 

\section*{Acknowledgements}
I thank the Centre for Climate Research Singapore (CCRS) for providing access to data from Singapore's Third National Climate Change Study (V3). Thanks also to my colleagues in the AI Downscaling team at the Met Office for helpful discussions on evaluation metrics, and in particular to James Redman for code reviews and discussions around flow matching. 

\bibliographystyle{plainnat}
\bibliography{references}

\begin{thebibliography}{24}
\providecommand{\natexlab}[1]{#1}
\providecommand{\url}[1]{\texttt{#1}}
\expandafter\ifx\csname urlstyle\endcsname\relax
  \providecommand{\doi}[1]{doi: #1}\else
  \providecommand{\doi}{doi: \begingroup \urlstyle{rm}\Url}\fi

\bibitem[Addison et~al.(2026)Addison, Kendon, Ravuri, Aitchison, and Watson]{cpmgem2024}
Henry Addison, Elizabeth Kendon, Suman Ravuri, Laurence Aitchison, and Peter~AG Watson.
\newblock Machine learning emulation of precipitation from km-scale uk regional climate simulations using a diffusion model, 2026.
\newblock URL \url{https://arxiv.org/abs/2407.14158}.

\bibitem[Albergo et~al.(2025)Albergo, Boffi, and Vanden-Eijnden]{fm_albergo}
Michael~S. Albergo, Nicholas~M. Boffi, and Eric Vanden-Eijnden.
\newblock Stochastic interpolants: A unifying framework for flows and diffusions, 2025.
\newblock URL \url{https://arxiv.org/abs/2303.08797}.

\bibitem[Ba{\~n}o-Medina et~al.(2022)Ba{\~n}o-Medina, Manzanas, Cimadevilla, Fern{\'a}ndez, Gonz{\'a}lez-Abad, Cofi{\~n}o, and Guti{\'e}rrez]{deepesd}
Jorge Ba{\~n}o-Medina, Rodrigo Manzanas, Ezequiel Cimadevilla, Jes{\'u}s Fern{\'a}ndez, Jose Gonz{\'a}lez-Abad, Antonio~Santiago Cofi{\~n}o, and Jos{\'e}~Manuel Guti{\'e}rrez.
\newblock Downscaling multi-model climate projection ensembles with deep learning (deepesd): contribution to cordex eur-44.
\newblock \emph{Geoscientific Model Development Discussions}, 2022:\penalty0 1--14, 2022.

\bibitem[{Centre for Climate Research Singapore}(2024)]{ccrs2024v3science}
{Centre for Climate Research Singapore}.
\newblock Singapore's third national climate change study: Science report.
\newblock Technical report, Meteorological Service Singapore, 2024.
\newblock URL \url{https://www.mss-int.sg/v3-climate-projections/resources/v3-reports}.
\newblock Available at \url{https://www.mss-int.sg/docs/default-source/v3_reports/v3_science_report/v3-science-report-full.pdf}.

\bibitem[Dhariwal and Nichol(2021)]{dhariwal2021diffusionmodelsbeatgans}
Prafulla Dhariwal and Alex Nichol.
\newblock Diffusion models beat gans on image synthesis, 2021.
\newblock URL \url{https://arxiv.org/abs/2105.05233}.

\bibitem[Dipankar et~al.(2020)Dipankar, Webster, Sun, Sanchez, North, Furtado, Wilkinson, Lock, Vosper, Huang, et~al.]{dipankar2020singv}
Anurag Dipankar, Stuart Webster, Xiangming Sun, Claudio Sanchez, Rachel North, Kalli Furtado, Jonathan Wilkinson, Adrian Lock, Simon Vosper, Xiang-Yu Huang, et~al.
\newblock Singv: A convective-scale weather forecast model for singapore.
\newblock \emph{Quarterly Journal of the Royal Meteorological Society}, 146\penalty0 (733):\penalty0 4131--4146, 2020.

\bibitem[Doury et~al.(2023)Doury, Somot, Gadat, Ribes, and Corre]{doury2023regional}
Antoine Doury, Samuel Somot, Sebastien Gadat, Aur{\'e}lien Ribes, and Lola Corre.
\newblock Regional climate model emulator based on deep learning: Concept and first evaluation of a novel hybrid downscaling approach.
\newblock \emph{Climate Dynamics}, 60\penalty0 (5):\penalty0 1751--1779, 2023.

\bibitem[Doury et~al.(2024)Doury, Somot, and Gadat]{doury2024suitability}
Antoine Doury, Samuel Somot, and Sebastien Gadat.
\newblock On the suitability of a convolutional neural network based rcm-emulator for fine spatio-temporal precipitation.
\newblock \emph{Climate Dynamics}, 62\penalty0 (9):\penalty0 8587--8613, 2024.

\bibitem[Fotiadis et~al.(2024)Fotiadis, Brenowitz, Geffner, Cohen, Pritchard, Vahdat, and Mardani]{sfm}
Stathi Fotiadis, Noah Brenowitz, Tomas Geffner, Yair Cohen, Michael Pritchard, Arash Vahdat, and Morteza Mardani.
\newblock Stochastic flow matching for resolving small-scale physics, 2024.
\newblock URL \url{https://arxiv.org/abs/2410.19814}.

\bibitem[Gutowski~Jr et~al.(2020)Gutowski~Jr, Ullrich, Hall, Leung, O’Brien, Patricola-DiRosario, Arritt, Bukovsky, Calvin, Feng, et~al.]{needforrcms}
William~J Gutowski~Jr, Paul~Aaron Ullrich, Alex Hall, L~Ruby Leung, Travis~Allen O’Brien, CM~Patricola-DiRosario, Raymond~W Arritt, Melissa~S Bukovsky, Katherine~V Calvin, Zhe Feng, et~al.
\newblock The ongoing need for high-resolution regional climate models: Process understanding and stakeholder information.
\newblock \emph{Bulletin of the American Meteorological Society}, 101\penalty0 (5):\penalty0 E664--E683, 2020.

\bibitem[Kendon et~al.(2012)Kendon, Roberts, Senior, and Roberts]{kendon2012realism}
Elizabeth~J Kendon, Nigel~M Roberts, Catherine~A Senior, and Malcolm~J Roberts.
\newblock Realism of rainfall in a very high-resolution regional climate model.
\newblock \emph{Journal of Climate}, 25\penalty0 (17):\penalty0 5791--5806, 2012.

\bibitem[Kendon et~al.(2025)Kendon, Addison, Doury, Somot, Watson, Booth, Coppola, Guti{\'e}rrez, Murphy, and Scullion]{kendon2025potential}
Elizabeth~J Kendon, Henry Addison, Antoine Doury, Samuel Somot, Peter~AG Watson, Ben~BB Booth, Erika Coppola, Jos{\'e}~Manuel Guti{\'e}rrez, James Murphy, and Calum Scullion.
\newblock Potential for machine learning emulators to augment regional climate simulations in provision of local climate change information.
\newblock \emph{Bulletin of the American Meteorological Society}, 106\penalty0 (6):\penalty0 E1175--E1203, 2025.

\bibitem[Leeuwenburg et~al.(2024)Leeuwenburg, Loveday, Ebert, Cook, Khanarmuei, Taggart, Ramanathan, Carroll, Chong, Griffiths, and Sharples]{Leeuwenburg_scores_A_Python_2024}
Tennessee Leeuwenburg, Nicholas Loveday, Elizabeth~E. Ebert, Harrison Cook, Mohammadreza Khanarmuei, Robert~J. Taggart, Nikeeth Ramanathan, Maree Carroll, Stephanie Chong, Aidan Griffiths, and John Sharples.
\newblock {scores: A Python package for verifying and evaluating models and predictions with xarray}.
\newblock \emph{Journal of Open Source Software}, 9\penalty0 (99):\penalty0 6889, July 2024.
\newblock \doi{10.21105/joss.06889}.
\newblock URL \url{https://joss.theoj.org/papers/10.21105/joss.06889}.

\bibitem[Leeuwenburg et~al.(2026)Leeuwenburg, Loveday, Ramanathan, Chong, Taggart, Shrestha, Khanarmuei, Cook, Bluett, Ebert, Carroll, Trotta, Sharples, Bishop, Squire, Griffiths, Pagano, Fisher, Mandelbaum, Jinghan, Smith, Abellan, Beunk, Esperson, Smallwood, and Wu]{leeuwenburg_2026_18638494}
Tennessee Leeuwenburg, Nicholas Loveday, Nikeeth Ramanathan, Stephanie Chong, Robert~J. Taggart, Durga Shrestha, Mohammadreza Khanarmuei, Harrison Cook, Liam Bluett, Elizabeth~E. Ebert, Maree Carroll, Belinda Trotta, John Sharples, Sam Bishop, Dougal~T. Squire, Aidan Griffiths, Thomas~C. Pagano, A.J. Fisher, Taylor Mandelbaum, Fu~Jinghan, Paul~R. Smith, Esteban Abellan, Jurian Beunk, Felix Esperson, J.~Smallwood, and Xiaoxi Wu.
\newblock scores: Metrics for the verification, evaluation and optimisation of forecasts, predictions or models, February 2026.
\newblock URL \url{https://doi.org/10.5281/zenodo.18638494}.

\bibitem[Lipman et~al.(2023)Lipman, Chen, Ben-Hamu, Nickel, and Le]{fm_lipman}
Yaron Lipman, Ricky T.~Q. Chen, Heli Ben-Hamu, Maximilian Nickel, and Matt Le.
\newblock Flow matching for generative modeling, 2023.
\newblock URL \url{https://arxiv.org/abs/2210.02747}.

\bibitem[Maraun et~al.(2010)Maraun, Wetterhall, Ireson, Chandler, Kendon, Widmann, Brienen, Rust, Sauter, Theme{\ss}l, et~al.]{maraun2010precipitation}
Douglas Maraun, Frederick Wetterhall, Anderson~M Ireson, Richard~E Chandler, Elizabeth~J Kendon, Martin Widmann, Stephan Brienen, Henning~W Rust, Tobias Sauter, Matthias Theme{\ss}l, et~al.
\newblock Precipitation downscaling under climate change: Recent developments to bridge the gap between dynamical models and the end user.
\newblock \emph{Reviews of geophysics}, 48\penalty0 (3), 2010.

\bibitem[Mardani et~al.(2024)Mardani, Brenowitz, Cohen, Pathak, Chen, Liu, Vahdat, Nabian, Ge, Subramaniam, Kashinath, Kautz, and Pritchard]{corrdiff}
Morteza Mardani, Noah Brenowitz, Yair Cohen, Jaideep Pathak, Chieh-Yu Chen, Cheng-Chin Liu, Arash Vahdat, Mohammad~Amin Nabian, Tao Ge, Akshay Subramaniam, Karthik Kashinath, Jan Kautz, and Mike Pritchard.
\newblock Residual corrective diffusion modeling for km-scale atmospheric downscaling, 2024.
\newblock URL \url{https://arxiv.org/abs/2309.15214}.

\bibitem[Pulkkinen et~al.(2019)Pulkkinen, Nerini, P\'erez~Hortal, Velasco-Forero, Seed, Germann, and Foresti]{pysteps}
S.~Pulkkinen, D.~Nerini, A.~A. P\'erez~Hortal, C.~Velasco-Forero, A.~Seed, U.~Germann, and L.~Foresti.
\newblock Pysteps: an open-source python library for probabilistic precipitation nowcasting (v1.0).
\newblock \emph{Geoscientific Model Development}, 12\penalty0 (10):\penalty0 4185--4219, 2019.
\newblock \doi{10.5194/gmd-12-4185-2019}.
\newblock URL \url{https://gmd.copernicus.org/articles/12/4185/2019/}.

\bibitem[Roberts and Lean(2008)]{fsslean2008}
Nigel~M. Roberts and Humphrey~W. Lean.
\newblock Scale-selective verification of rainfall accumulations from high-resolution forecasts of convective events.
\newblock \emph{Monthly Weather Review}, 136\penalty0 (1):\penalty0 78 -- 97, 2008.
\newblock \doi{10.1175/2007MWR2123.1}.
\newblock URL \url{https://journals.ametsoc.org/view/journals/mwre/136/1/2007mwr2123.1.xml}.

\bibitem[Song et~al.(2021)Song, Sohl-Dickstein, Kingma, Kumar, Ermon, and Poole]{song2021scorebasedgenerativemodelingstochastic}
Yang Song, Jascha Sohl-Dickstein, Diederik~P. Kingma, Abhishek Kumar, Stefano Ermon, and Ben Poole.
\newblock Score-based generative modeling through stochastic differential equations, 2021.
\newblock URL \url{https://arxiv.org/abs/2011.13456}.

\bibitem[Tancik et~al.(2020)Tancik, Srinivasan, Mildenhall, Fridovich-Keil, Raghavan, Singhal, Ramamoorthi, Barron, and Ng]{tancik2020fourierfeaturesletnetworks}
Matthew Tancik, Pratul~P. Srinivasan, Ben Mildenhall, Sara Fridovich-Keil, Nithin Raghavan, Utkarsh Singhal, Ravi Ramamoorthi, Jonathan~T. Barron, and Ren Ng.
\newblock Fourier features let networks learn high frequency functions in low dimensional domains, 2020.
\newblock URL \url{https://arxiv.org/abs/2006.10739}.

\bibitem[Van Der~Meer et~al.(2023)Van Der~Meer, de~Roda~Husman, and Lhermitte]{van2023deep}
Marijn Van Der~Meer, Sophie de~Roda~Husman, and Stef Lhermitte.
\newblock Deep learning regional climate model emulators: A comparison of two downscaling training frameworks.
\newblock \emph{Journal of Advances in Modeling Earth Systems}, 15\penalty0 (6):\penalty0 e2022MS003593, 2023.

\bibitem[Ward-Leikis et~al.(2025)Ward-Leikis, Rampal, Koh, Gibson, Liu, Kitsios, Meyers, Adie, Juntao, and Sherwood]{wardleikis2025intercomparisongenerativemachinelearning}
Bryn Ward-Leikis, Neelesh Rampal, Yun~Sing Koh, Peter~B. Gibson, Hong-Yang Liu, Vassili Kitsios, Tristan Meyers, Jeff Adie, Yang Juntao, and Steven~C. Sherwood.
\newblock An intercomparison of generative machine learning methods for downscaling precipitation at fine spatial scales, 2025.
\newblock URL \url{https://arxiv.org/abs/2512.13987}.

\bibitem[Wernli et~al.(2008)Wernli, Paulat, Hagen, and Frei]{wernlisal2008}
Heini Wernli, Marcus Paulat, Martin Hagen, and Christoph Frei.
\newblock Sal—a novel quality measure for the verification of quantitative precipitation forecasts.
\newblock \emph{Monthly Weather Review}, 136\penalty0 (11):\penalty0 4470 -- 4487, 2008.
\newblock \doi{10.1175/2008MWR2415.1}.
\newblock URL \url{https://journals.ametsoc.org/view/journals/mwre/136/11/2008mwr2415.1.xml}.

\end{thebibliography}

\end{document}